\documentclass[a4paper,10pt]{article}
\usepackage{amsmath}
\usepackage{parskip}
\usepackage{ucs}

\usepackage[utf8]{inputenc} 
\usepackage[pdftex]{graphicx}

\usepackage{subcaption}
\graphicspath{{./Pic/}}
\DeclareGraphicsExtensions{.pdf,.png,.jpg}
\usepackage{ulem}
\usepackage{hhline}
\usepackage{color,colortbl}
\usepackage{authblk}

\title{Impact of the improved parallel kinetic coefficients on the helium and neon transport in SOLPS-ITER for ITER}
\author[1]{S.~Makarov}
\author[1]{D.~Coster}
\author[2]{V.~Rozhansky}
\author[2]{S.~P.~Voskoboynikov}
\author[2]{E.~Kaveeva}
\author[2]{I.~Senichenkov}
\author[3]{A.~Stepanenko}
\author[3]{V.~Zhdanov}
\author[4]{X.~Bonnin}

\affil[1]{Max-Planck-Institut f\"ur Plasmaphysik, D-85748 Garching, Germany}
\affil[2]{Peter the Great St.Petersburg Polytechnic University, St.Petersburg, Russia}
\affil[3]{National Research Nuclear University MEPhI (Moscow Engineering Physics Institute), Kashirskoe sh. 31,
115409, Moscow, Russia}
\affil[4]{ITER Organization, CS 90 046, F-13067, St-Paul-Lez-Durance Cedex, France}

\begin{document}

\maketitle

\begin{abstract}
{New Grad's-Zhdanov module is implemented in the SOLPS-ITER code and applied to ITER impurity transport simulations. Significant difference appears in the helium transport due to improved parallel kinetic coefficients. As a result 30\% decrease of the separatrix-averaged helium relative concentration is observed for the constant helium source and pumping speed. Change of the impurity behaviour is discussed. For the neon changes are less pronounced. For the first time the ion distribution functions are studied in the ITER Scrape-off layer conditions to reveal the origin of the kinetic coefficient improvements and theory limitations.}
\end{abstract}

\section{Introduction}
Impurity transport is a crucial problem in fusion research. Impurity accumulation in the confinement region of fusion devices should be avoided. Edge and Scrape-off layer (SOL) transport affects significantly global impurity transport. The SOLPS-ITER code has been developed to study transport in the SOL and Edge \cite{schneider2006plasma,WIESEN2015480}. In the standard SOLPS-ITER model Zhdanov-Yushmanov expressions \cite{book} are used for the thermal and friction forces coefficients calculation \cite{Sytova2020}. However, for the ions with close masses the complete Grad's-Zhdanov method should be used \cite{book}. It is important to mention that multi-temperature generalisation of this method is in progress for the Soledge3x-EIRENE code \cite{Raghunathan2021,Raghunathan2021_2}. In the present paper a new Grad's-Zhdanov module based on the improved approach \cite{Makarov2021} and complete Grad's-Zhdanov method \cite{book,Makarov2021} has been implemented in the SOLPS-ITER code (Sec. \ref{module_sec}). Impact on the helium and neon transport is studied (Sec. \ref{impurity_sec}). The new approach leads to changes in the parallel impurity momentum balance, which causes the poloidal impurity stagnation point to shift in the divertor region. The solution is very sensitive to the stagnation point variations. A simple model is proposed to explain how small changes of the stagnation point lead to the significant difference in the impurity balance. For the first time for the ITER SOL, the ion distribution functions are studied using this method to examine the origin of the kinetic coefficients changes (Sec. \ref{dist_sec}). The applicability of the method is also discussed.
\section{Grad's-Zhdanov module}\label{module_sec}
In the paper \cite{Makarov2021} the reduced form of the improved analytical expressions was implemented into the SOLPS-ITER code and tested for the neon transport in ITER. Now a complete Grad's-Zhdanov module is implemented into the SOLPS-ITER code for the kinetic coefficients calculation using the 21N-moment method \cite{book}. This module operates in two regimes: 1) Improved analytical method (IAM) \cite{Makarov2021}; 2) Explicit matrix inversion method (EMIM) \cite{book,Makarov2021}. The new module calculates thermal and friction forces coefficients between all ions, whereas the standard approach considered the thermal force only due to main ion interaction with impurities. Moreover, the heat flux according to the Grad's-Zhdanov closure is applied instead of the Braginskii ion heat conductivity \cite{braginskii1965transport}. Also, the ion velocity difference dependent heat flux was taken into account. Even though no significant impact from the new heat flux was observed on the solution, now all vector parallel moments are considered according to the 21N-moment Grad's-Zhdanov method. Therefore, the thermal and friction forces are treated self-consistently with the heat flux. 
%Important to mention that application of the numerical method does not lead to the significant performance reduction of the SOLPS-ITER code, however application of the analytical expressions makes further analysis easier in case where both approaches provide similar results. 
The Grad's-Zhdanov stress-viscosity implementation is underway. 
\section{Impurity transport simulation results in ITER and sensitivity tests}\label{impurity_sec}
\begin{figure*}[htbp]
    \begin{subfigure}{0.3\textwidth}
        \includegraphics[scale=0.12]{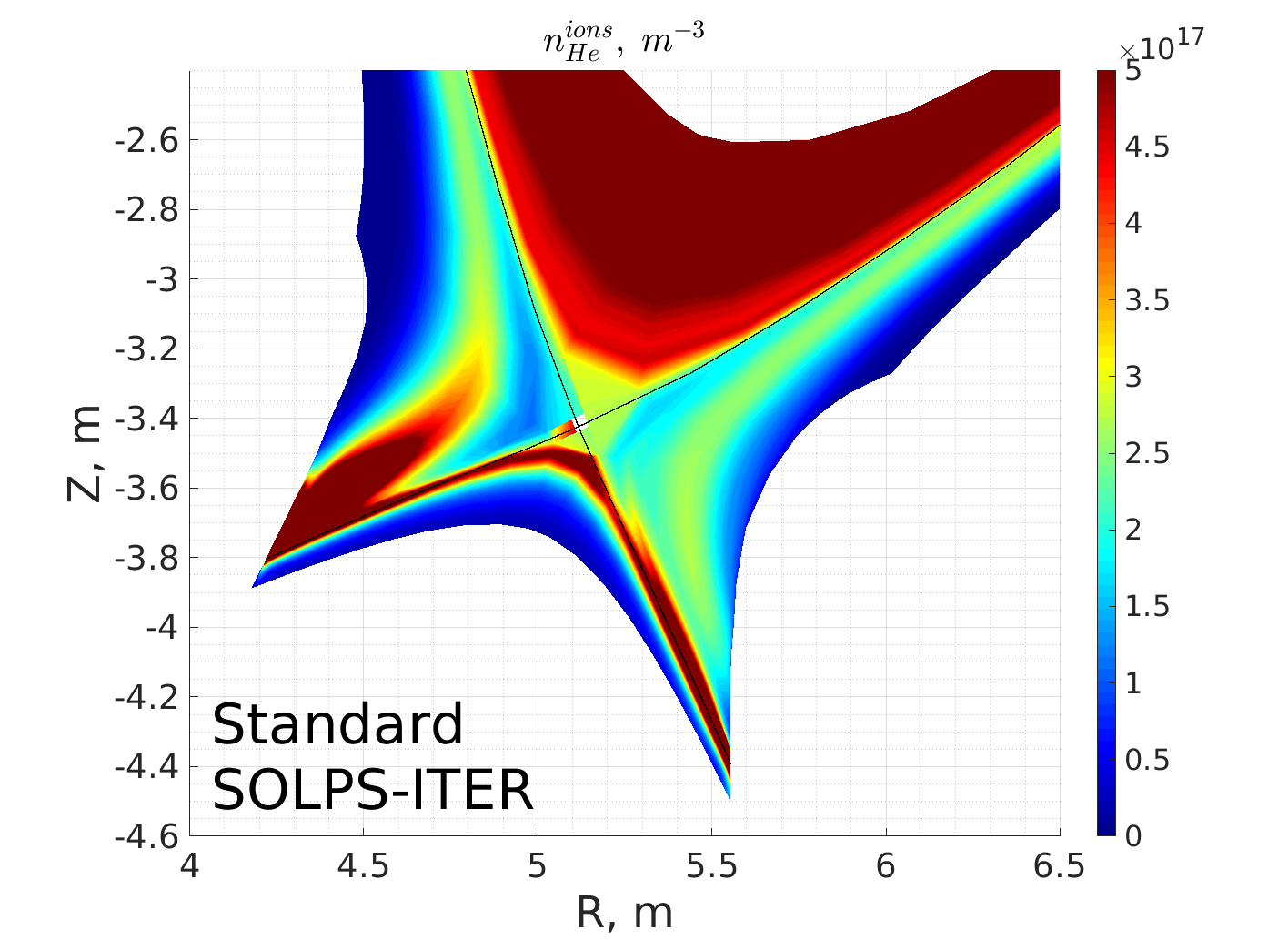}  
        \caption{}
        \label{fig:Iter_nNe_Original}
    \end{subfigure}
    \begin{subfigure}{0.3\textwidth}
        \includegraphics[scale=0.12]{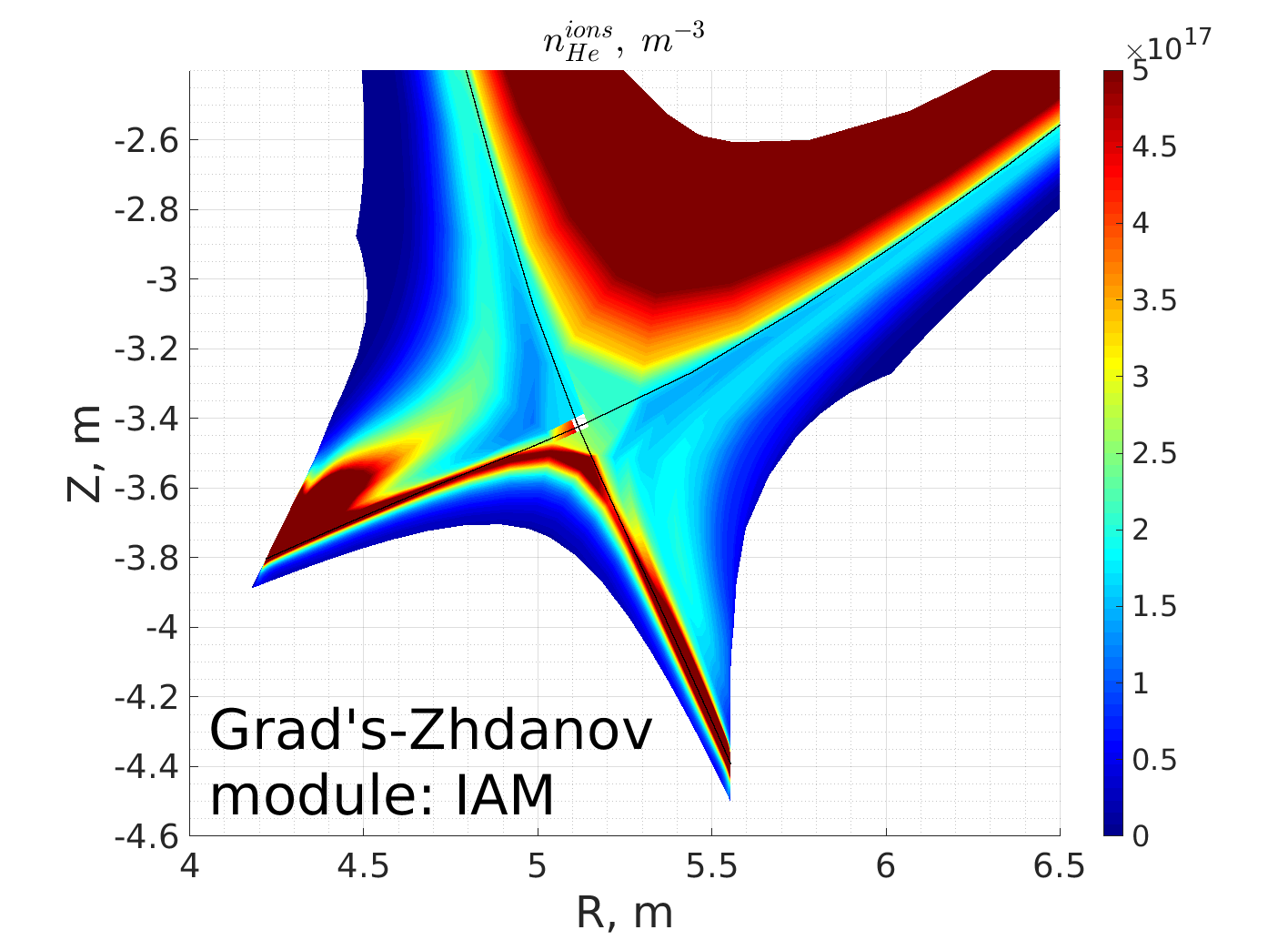}  
        \caption{}
    
        \label{fig:Iter_nNe_Imrpoved}
    \end{subfigure}
        \begin{subfigure}{0.3\textwidth}
        \includegraphics[scale=0.12]{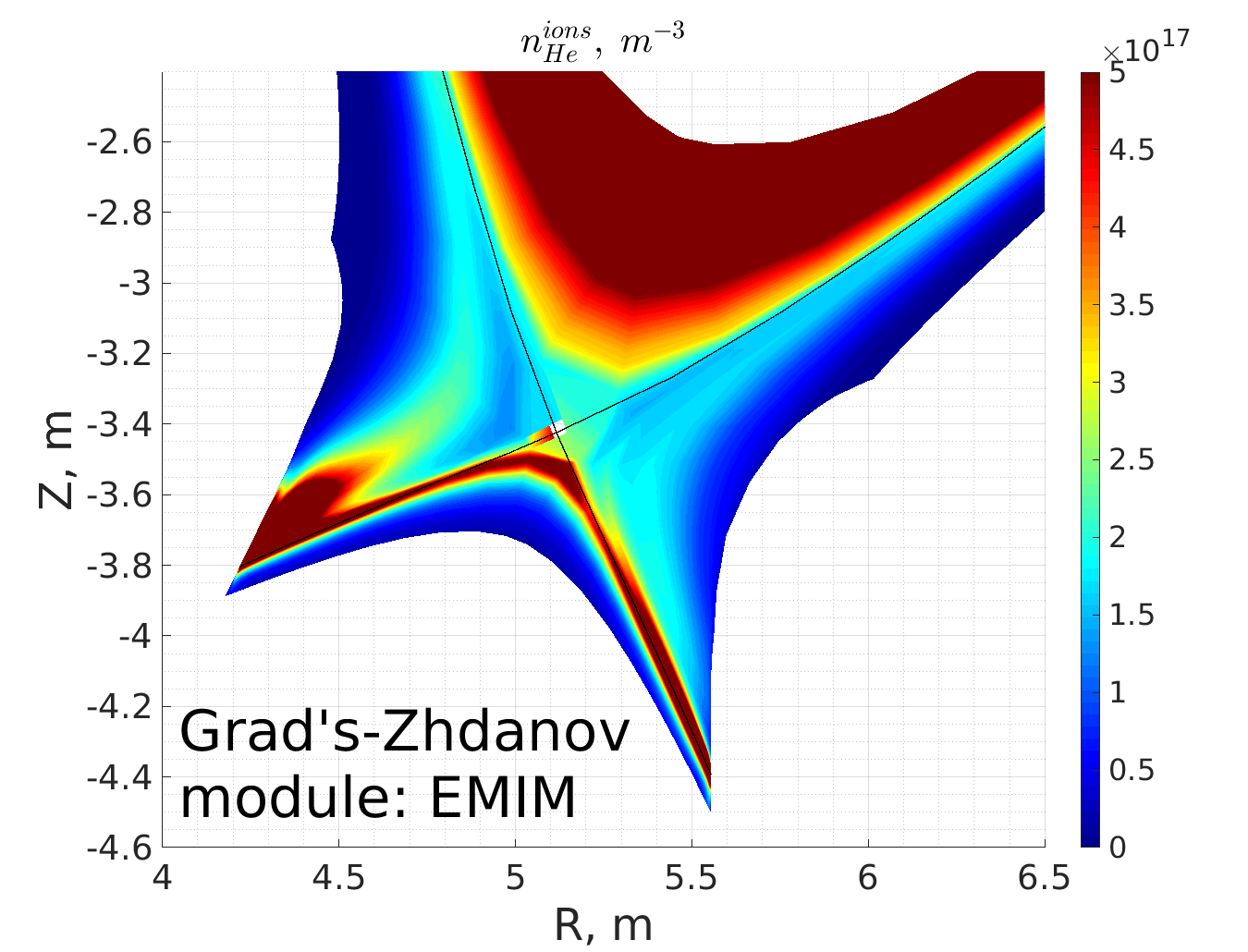}  
        \caption{}
        \label{fig:Iter_nNe_Imrpoved}
    \end{subfigure}
    \caption{Helium ion density in the divertor region for the Standard SOLPS-ITER model (a) and the Grad's-Zhdanov module applied: IAM (b) and EMIM (c)}\label{fig:nHe_div}
\end{figure*}
In this section, helium and neon transport for 500MW baseline ITER scenario (Energy flux through the separatrix $P_{SOL}\approx 100MW$) using the new Grad's-Zhdanov module is considered. The simulation has been done for the deuterium, helium and neon mixture with full drifts and currents and coupled with EIRENE. %\cite{Reiter2005}.
Investigation of the case, which  has been presented in our previous paper \cite{Makarov2021}, was continued. It is catalogued in the ITER integrated modeling analysis suite (IMAS) database as 123081. 
%Deuterium thoughput and pumping speed are kept the same, and as a result the divertor neutral pressure is $p_n=7.5Pa$.

%Thus, all three approaches are compared, such as: Zhdanov-Yushmanov (ZY) original analytical expressions (8.4.7) \cite{book}, our improved analytical expressions from \cite{Makarov2021} and explicit numerical matrix inversion. As it showed in \cite{Makarov2021} the relative separatrix-averaged Ne concentration dropped to $c_{Ne}=0.8\%$ after application of either our improved analytical expressions or numerical matrix inversion. Significant difference in the neon transport between our improved analytical expression and the matrix inversion method is not observed, as expected \cite{Makarov2021}. Helium transport is more affected by using our new analytical expressions and the direct numerical inversion, due to closer mass between helium and deuterium. 
First of all, we consider helium transport. After the implementation of the new module, the separatrix-averaged helium concentration has dropped from $c_{He}=0.86\%$ for the standard approach (close to results in \cite{KUKUSHKIN20112865}) to $c_{He}=0.61\%$ for the analytical (IAM) and $c_{He}=0.57\%$ for the numerical (EMIM) mode correspondingly. Also, one can see the difference of the helium distribution in the divertor and SOL (figure \ref{fig:nHe_div}). For the new approach helium is better compressed in the divertor. In figure \ref{fig:nHe_div} it can be seen that the helium density above the X-point is lower for this method than for the standard one. Moreover, the high helium density region is pushed closer to the target in the inner divertor (figure \ref{fig:nHe_div}). 
%This effect can be explained according to the analysis made previously \cite{Makarov2021}. 

\begin{figure*}[htbp]
    \begin{subfigure}{0.3\textwidth}
        \includegraphics[scale=0.2]{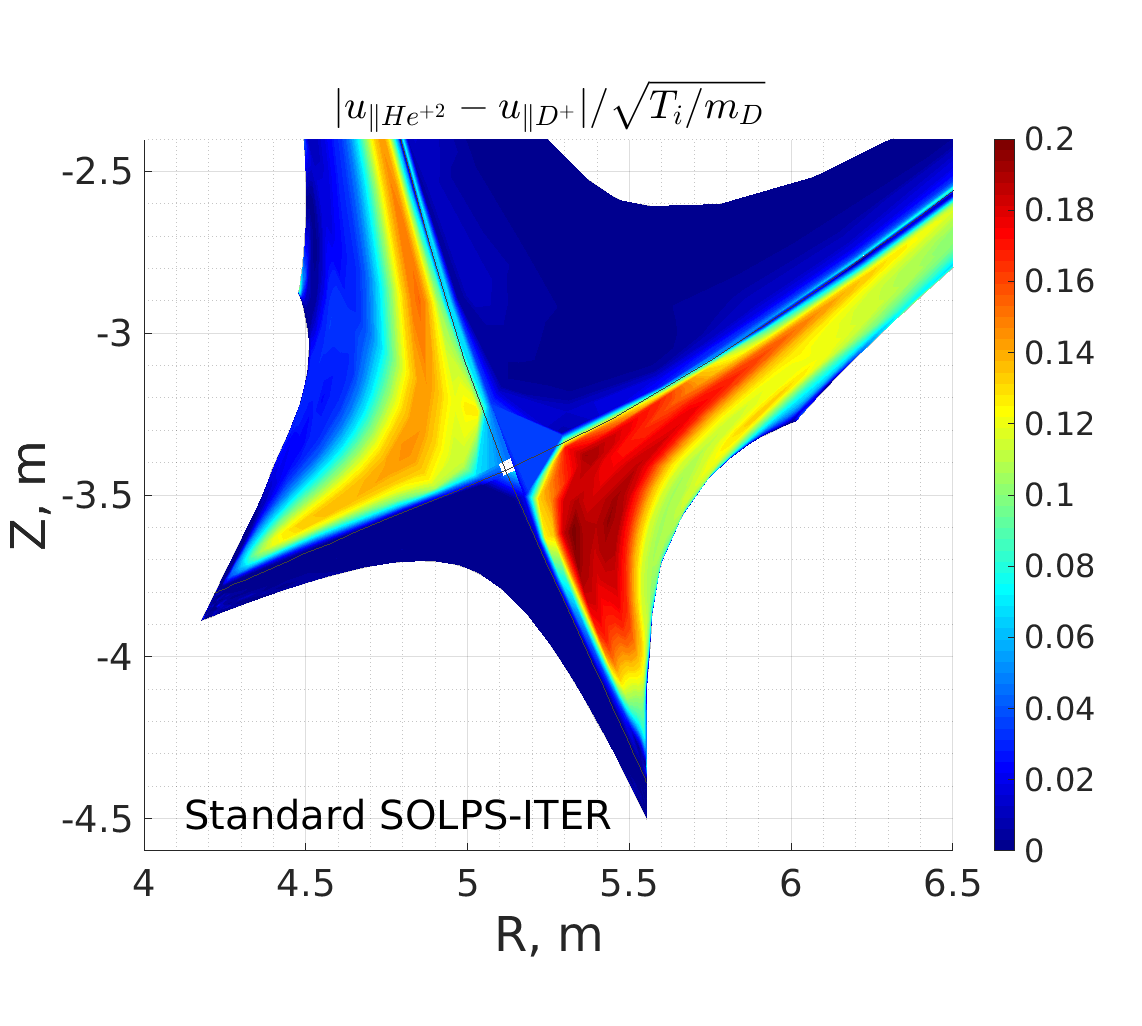}  
        \caption{}
        \label{fig:He_vel_diff_Original}
    \end{subfigure}
     \qquad \qquad \qquad \qquad
    \begin{subfigure}{0.3\textwidth}
        \includegraphics[scale=0.2]{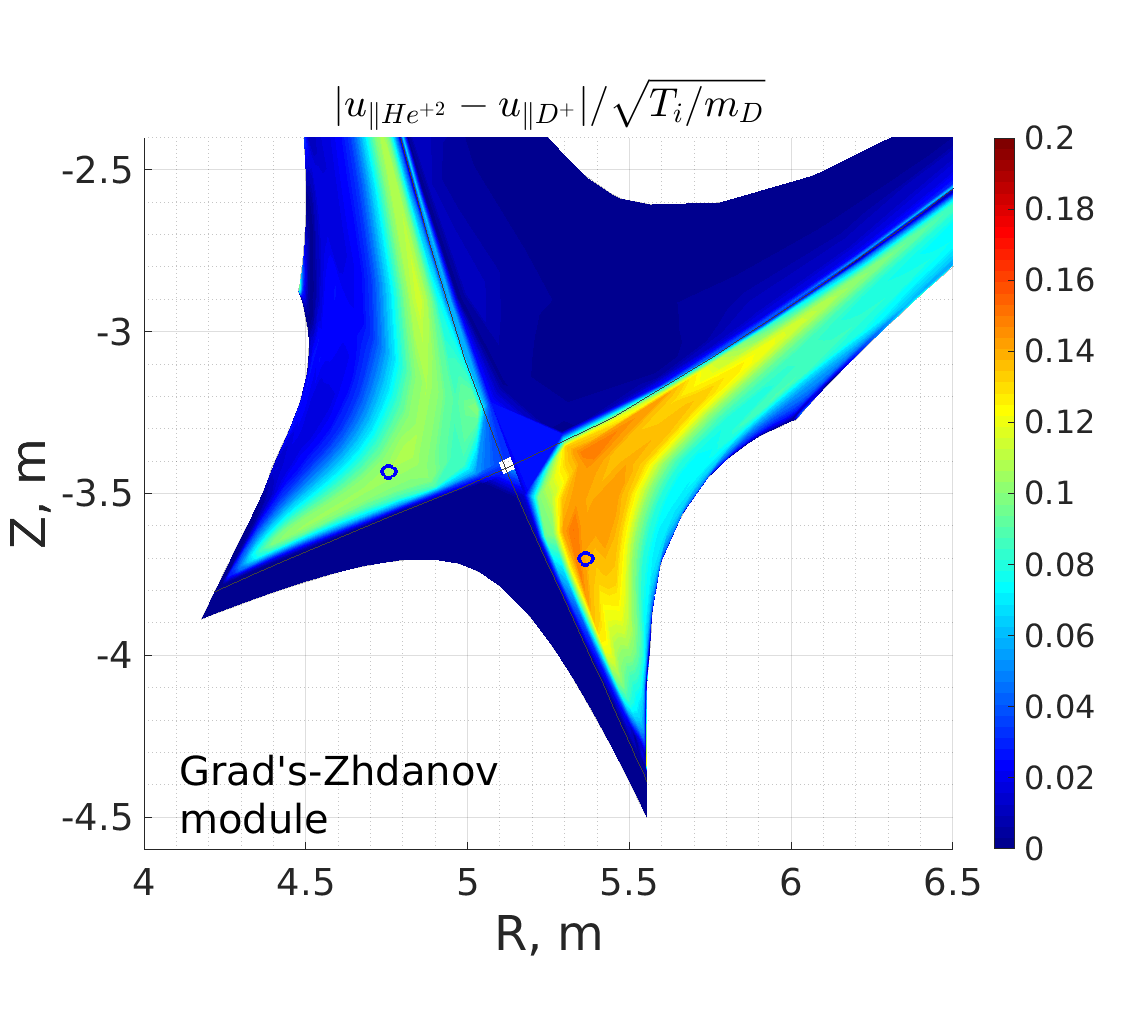}  
        \caption{}
        \label{fig:He_vel_diff_Imrpoved}
    \end{subfigure}
    \caption{The helium ion velocity relative to the deuterium ion velocity and normalised to the deuterium thermal velocity for the standard SOLPS-ITER model (a) and the Grad's-Zhdanov module application (IAM) (b). Blue circles define spatial points where ion distribution functions are studied in section \ref{dist_sec}.}\label{fig:He_vel_diff}
\end{figure*}

\begin{figure*}[htbp]
\includegraphics[scale=0.3]{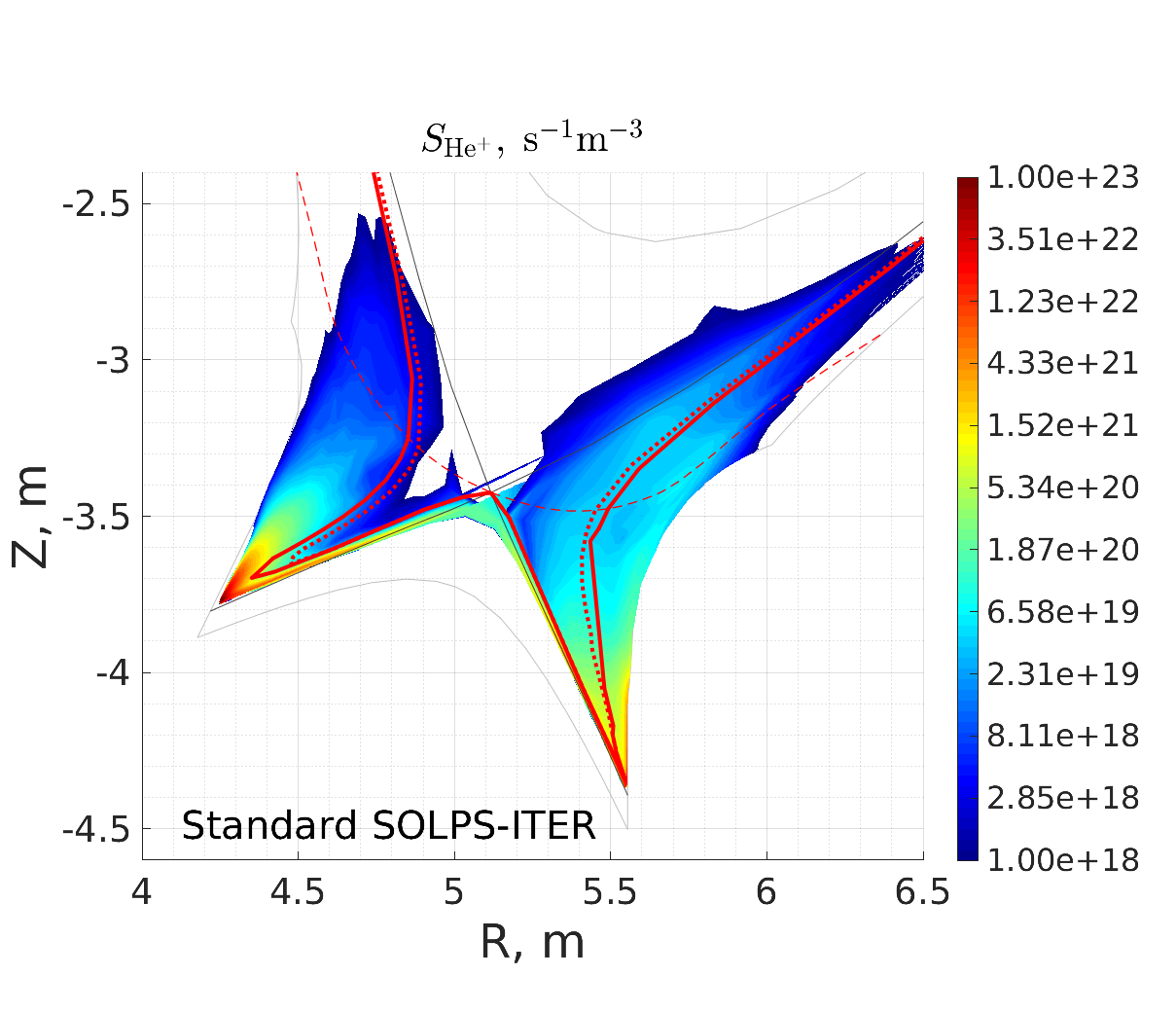}  
    \caption{The helium ionisation/recombination source for the standard model. Red solid line connects helium poloidal stagnation points for the standard approach; red dotted line connects helium poloidal stagnation points after the Grad's-Zhdanov module application (IAM). Red dashed line - the divertor region boundry. }\label{fig:Original_SHe_div}
\end{figure*}

Indeed, for the helium the new approach provides larger kinetic coefficient for the friction force and smaller kinetic coefficient for the thermal force than the standard SOLPS-ITER approach. In other words, the helium thermal diffusion (balance between friction and thermal forces), which drags helium upstream, %and prevents helium to be efficiently pumped out
is weaker for given parallel temperature gradient for the improved method. It can be illustrated in terms of impurity-deuterium velocity difference, which in the Scrape-off layer is mostly determined by the interplay between thermal and friction forces of impurity with deuterium  \cite{Senichenkov_2019}. Since the difference between analytical and numerical methods is not significant for helium, we will further focus on the comparison between Grad's-Zhdanov module application using analytical method and the standard SOLPS-ITER model. In figure \ref{fig:He_vel_diff} one can see the velocity difference between helium and deuterium normalised by local deuterium thermal velocity. 
%In the inner divertor helium-deuterium flow velocity difference $u_{\parallel He^{+2}}-u_{\parallel D^+}$ is positive, and in the outer divertor it is negative (parallel axis starts on the inner target and goes along the magnetic field towards the outer target).Therefore,
In the regions where the deuterium flow velocity is directed towards the target and the ion temperature gradient is sufficiently high, the helium velocity is directed from the divertor towards upstream. This shifts the helium stagnation point closer to the target. Its relative position to the impurity ionisation source is one of the major effects that defines impurity leakage from the divertor \cite{Senichenkov_2019}. The neutrals, which are ionized above the stagnation point, leak from the divertor and contribute significantly in the ion leakage flow. Indeed, for the standard approach the helium ionisation source further upstream than the stagnation point in each flux tube in the divertor region is $9.4\cdot 10^{20}\ particles/s$ (total for both inner and outer targets), whereas the helium ion leakage upstream from the divertor region is $10.5\cdot 10^{20}\ particles/s$ (the difference is due to a small radial ion flow inside the stagnation points region). After the Grad's-Zhdanov module application the helium stagnation point is shifted further upstream (figure \ref{fig:Original_SHe_div}), and the ionisation source further upstream than the stagnation point is $5.9\cdot 10^{20}\ particles/s$, whereas the helium ion leakage upstream from the divertor region is $7.2\cdot 10^{20}\ particles/s$. However, the significant change of the stagnation point appears only in the inner divertor, whereas a much larger contribution in the helium leakage comes from the outer divertor, where the stagnation point does not change significantly (figure \ref{fig:Original_SHe_div}).

\begin{figure*}[htbp]
\includegraphics[scale=0.3]{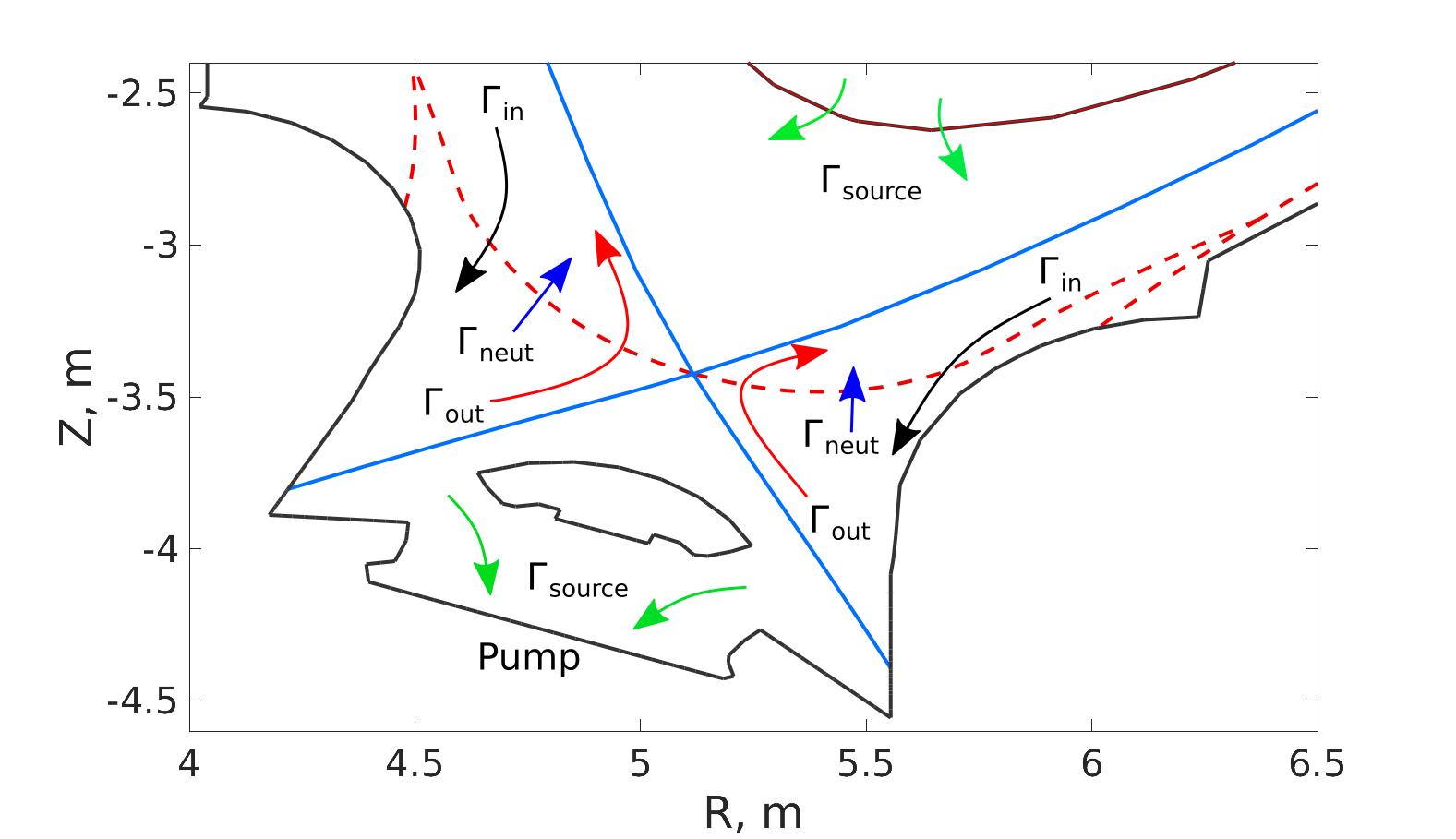}  
    \caption{Scheme of the main helium flows, which enter and leave the divertor region. Red dashed line - the divertor region boundry. Entering impurity flow splits into ion and neutral leakage flows and pumped out flow.}\label{fig:New_sketch}
\end{figure*}

To give the reader an understanding why small changes in the stagnation point significantly affect particle flows, a simple model is proposed:
\begin{align}\label{eq:share}
\Gamma_{out}=\alpha\cdot\Gamma_{in}; \ \ \ 
\Gamma_{neut}=\beta\cdot\Gamma_{in}
\end{align}
\begin{align}\label{eq:global_balance}
\Gamma_{in}=\Gamma_{out}+\Gamma_{neut}+\Gamma_{source}
\end{align}
where $\Gamma_{in}$ is the entering impurity ion poloidal flow into the divertor region through the Far-SOL; $\Gamma_{out}$ is the impurity ion poloidal leakage flow from the divertor region closer to the separatrix, and $\alpha$ is the fraction of entering ion poloidal flow which leaves the divertor region as ions; $\Gamma_{neut}$ is the impurity neutral leakage flow from the divertor region, which penetrates into the SOL above X-point and into the confinement region, and $\beta$ is the fraction of entering ion poloidal flow, which leaves the divertor region as neutrals; $\Gamma_{source}$ is the flow due to the imprity source (figure \ref{fig:New_sketch}). Note, the exact choice of the divertor region boundary (red dash in the figure \ref{fig:Original_SHe_div} and \ref{fig:New_sketch}) is arbitrary, however it only slightly affects distribution between $\alpha$ and $\beta$.  Eq. \eqref{eq:global_balance} is a global particle balance. As a result:
\begin{align}\label{eq:result}
\Gamma_{out}=\frac{\alpha}{1-\alpha-\beta}\Gamma_{source}
\end{align}
Note, fractions of the entering impurity flow $\alpha$ and $\beta$ are measured directly according to the modeling results. 
%In these simulations the fraction of the neutral leakage is the same: $\beta=0.18$.
The ion leakage fraction is $\alpha=0.76$ and $\beta=0.18$ for the standard approach and $\alpha=0.74$ and $\beta=0.16$ for our improved approach. The sum $\alpha+\beta$ is equal to  0.94\ (or 0.90 for the new module). It makes the denominator in \eqref{eq:result} small and the result \eqref{eq:result} sensitive to the denominator variation. Therefore, for chosen helium source (close to the fusion ash production) $\Gamma_{source}=1.0\cdot 10^{20}\ particles/s$, the ion helium leakage is $\Gamma_{out}=12.7\cdot 10^{20}\ particles/s$ and $\Gamma_{out}=7.4\cdot 10^{20}\ particles/s$ for the standard and improved (IAM) approach correspondingly. It is close to the results of the simulation, which is discussed earlier. Thus, a 4\% change of the distribution of the entering flow leads to a $>$30\% decrease of the ion leakage flow. This is the result of the feedback response of the entrance ion flow, which is described by Eq \eqref{eq:share}-\eqref{eq:global_balance}. Indeed, the less flow leaks from the divertor, the less flow returns back into the divertor and this affects the leakage flow, as a result. Similar feedback occur in the recycling phenomena on the target. This significant reduction of the impurity leakage flow then leads to the separatrix-averaged relative helium concentration drop. Finally it is worth to mention that this change of $\alpha$ is a result of the stagnation point shifting. Indeed, the fraction of the entering impurity flow, which penetrates above the stagnation point as neutrals and then got ionized there, is 69\%. If the stagnation point boundary is simply shifted from the standard case (red solid line in figure \ref{fig:Original_SHe_div}) to the case with Grad’s-Zhdanov module applied (red dotted line in figure \ref{fig:Original_SHe_div}) without changing the ionisation source profile, the ionisation source above the new boundary becomes $1.2\cdot 10^{20}\ particles/s$ less or 8\% in terms of the entering flow fraction.

\begin{figure*}[htbp]
    \begin{subfigure}{0.3\textwidth}
        \includegraphics[scale=0.2]{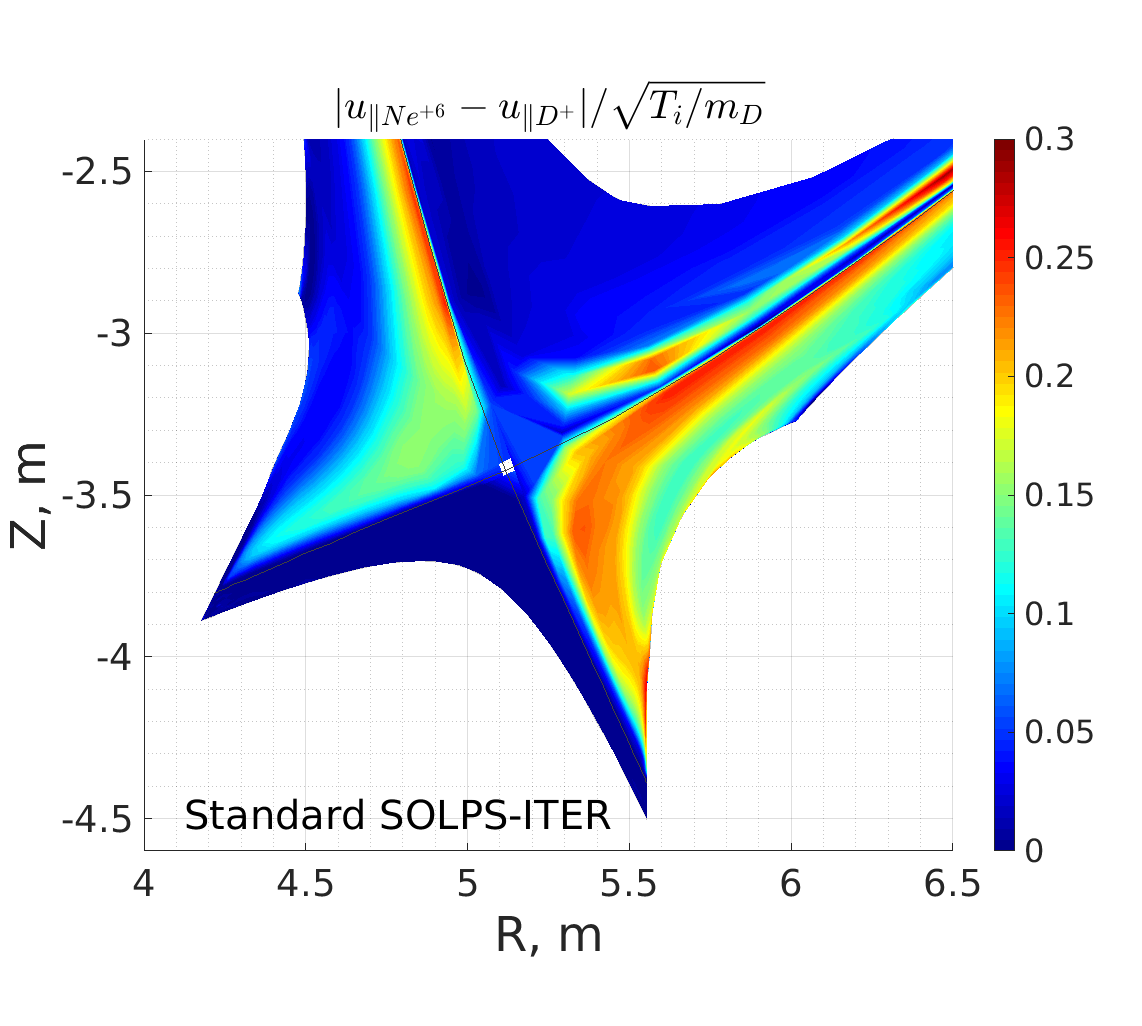}  
        \caption{}
        \label{fig:Ne_vel_diff_Original}
    \end{subfigure}
     \qquad \qquad \qquad \qquad
    \begin{subfigure}{0.3\textwidth}
        \includegraphics[scale=0.2]{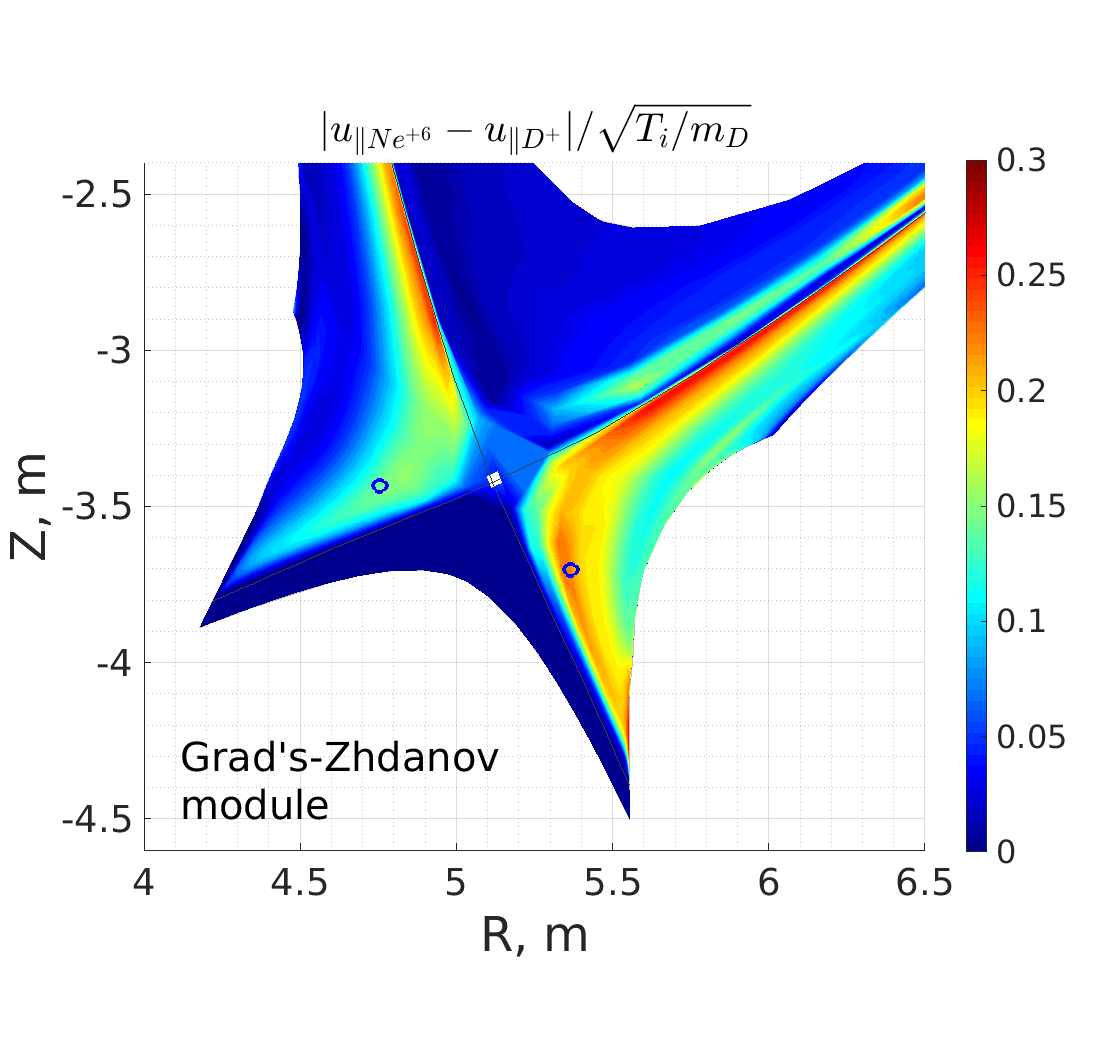}  
        \caption{}
        \label{fig:Ne_vel_diff_Imrpoved}
    \end{subfigure}
    \caption{The neon ion velocity relative to the deuterium ion velocity and normalised to the deuterium thermal velocity for the Standard SOLPS-ITER model (a) and the Grad's-Zhdanov module application (IAM) (b). Blue circles define spatial points, where ion distribution functions are studied in section \ref{dist_sec}.}\label{fig:Ne_vel_diff}
\end{figure*}

The same effect, however less pronounced, takes place for neon, as well (figure  \ref{fig:Ne_vel_diff}) (here the source is a seeding valve above the mid-plane). Before application of the Grad's-Zhdanov module the  separatrix-averaged relative neon concentration is $c_{Ne}=1.0\%$\footnote{The case (123081) was re-simulated using the latest to date (07.01.2021) 3.0.7 SOLPS-ITER version. The updated version caused decreasing of the neon presence in the plasma volume for the constant seeding rate. In this paper the reason of this drop is not discussed. However, to keep the separatrix-averaged relative neon concentration unchanged, the neon seeding rate from $0.6\cdot 10^{20}\ particles/s$ to $1.0\cdot 10^{20}\ particles/s$ was increased.}. After application of the Grad's-Zhdanov module it drops to $c_{Ne}=0.8\%$. Indeed, a smaller (than for the standard method) fraction of the impurity flow returns upstream and a larger fraction of the impurity flow goes into the pumping system, as it was mentioned in our earlier work \cite{Makarov2021}.

It is also worth to mention that in such ITER baseline scenario simulations the velocity difference for the helium can be up to 20\% of the deuterium thermal velocity  (figure \ref{fig:He_vel_diff}), and for the neon  up to 25\% of the deuterium thermal velocity (figure \ref{fig:Ne_vel_diff}) . This means, we are close to the limit of applicability of the linear collisional theory \cite{book}. This will be discussed in more detail in section \ref{dist_sec}.

\section{Ion distribution functions}\label{dist_sec}
\begin{figure*}[htbp]
    \begin{subfigure}{0.45\textwidth}
        \includegraphics[scale=0.15]{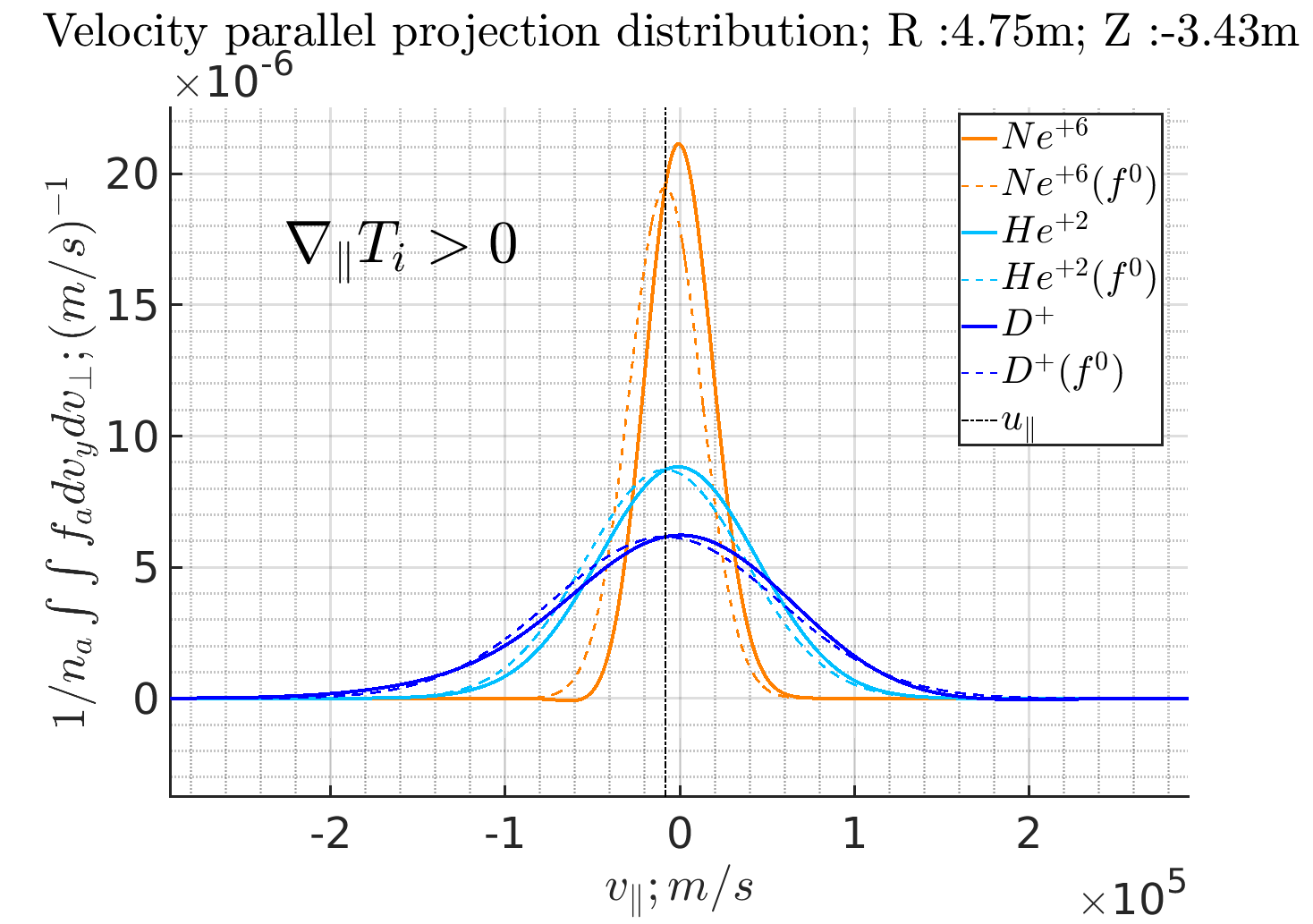}  
        \caption{}
    \end{subfigure}
    \qquad
    \begin{subfigure}{0.45\textwidth}
        \includegraphics[scale=0.15]{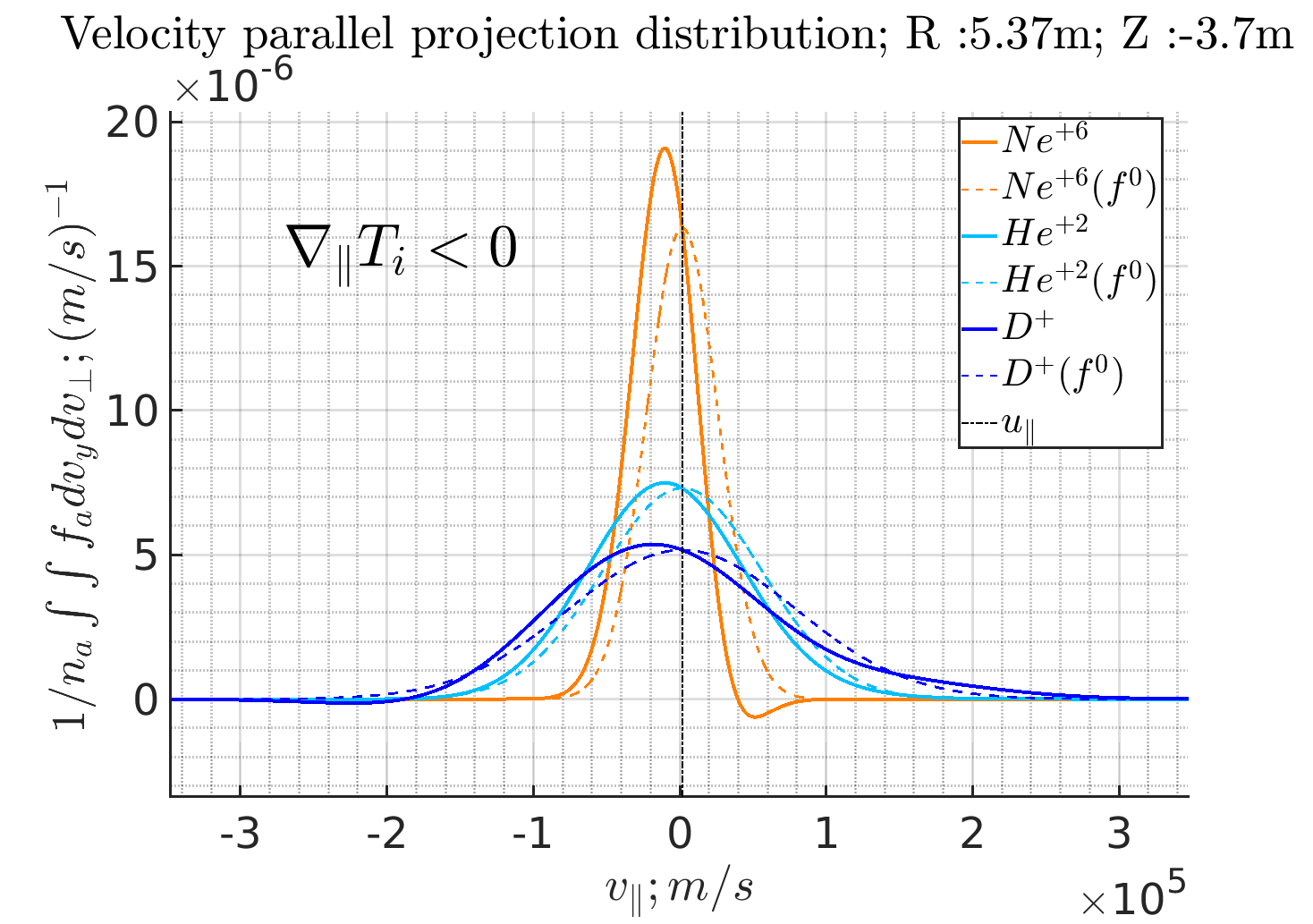}  
        \caption{}
        \label{fig:dist_fun_outer}
    \end{subfigure}
    \caption{Distribution functions of different ion species in the (a) inner and (b) outer divertor regions (at the spatial points marked by blue circles in figures \ref{fig:He_vel_diff_Imrpoved} and \ref{fig:Ne_vel_diff_Imrpoved}).}
        \label{fig:Iter_nNe_Original}\label{fig:dist_fun}
\end{figure*}

In this section, ion distribution functions for deuterium, helium and neon are presented for the ITER simulation with the aim of showing why the helium kinetic coefficients are affected more than the neon ones and discussing the method applicability. The ion distribution function, which is used to obtain kinetic coefficients, can be reconstructed based on its moments (they are calculated in the code). Indeed, let us consider the Hermite polynomial approximation of the distribution function that is used in this method \cite{book}: 
\begin{multline}
f_a(\textbf{c})=n_a\left(\frac{m_a}{2\pi T_i}\right)^{3/2}\exp{\left(-\frac{m_ac^2}{2T_i}\right)}\cdot\Bigg[1+\gamma_a w_{\parallel a}c_\parallel+\\ 
\frac{1}{5}\gamma_a^2\left(\frac{h^T_{\parallel a}}{p_a}+\frac{h^w_{\parallel a}}{p_a}\right)c_\parallel\left(c^2-\frac{5}{\gamma_a}\right)+\frac{1}{70}\gamma_a^3\left(\frac{r^T_{\parallel a}\gamma_a}{p_a}+\frac{r^w_{\parallel a}\gamma_a}{p_a}\right)c_\parallel\left(c^4-\frac{14}{\gamma_a}c^2+\frac{35}{\gamma_a^2}\right)\Bigg]
\end{multline}
where moments, which we output from the code, are: $w_{\parallel a}$ is the diffusive velocity; $h^T_{\parallel a}/ h^w_{\parallel a}$ and $r^T_{\parallel a}/r^w_{\parallel a}$ are the temperature gradient/velocity difference dependent parts of the heat flux and additional vector moment correspondingly (for details see \cite{book}). And:
\begin{align}
\gamma_a=\frac{m_a}{T_i},\ \ \ \textbf{c}=\textbf{v}-\textbf{u},\ \ \ \textbf{u} = \frac{\sum_{a} m_a n_{a} \textbf{u}_a}{\sum_{a} m_a n_{a}},\ \ \  \textbf w_{a}=\textbf u_{a}-\textbf u
\end{align}
Two spatial points are chosen: one in the inner divertor and one in the outer divertor (blue circles in figures \ref{fig:He_vel_diff_Imrpoved}, \ref{fig:Ne_vel_diff_Imrpoved}). In figure \ref{fig:dist_fun} predicted ion distribution functions at these points are shown. The parallel momentum transfer is studied, thus distribution functions are integrated over the perpendicular velocities. 

One of the main improvements for the thermal and friction forces calculation is the inclusion of the correction for the impurity distribution function due to the impact from the impurity heat flux and impurity additional vector moment \textit{r}, whereas for the standard SOLPS-ITER model only light main ion heat flux and additional vector moment \textit{r} affect the momentum collision term. The shaping of the impurity distribution function becomes important, when the $m_{D}/m_{impurity}$ ratio is not small, since the widths of distribution functions become comparable (figures \ref{fig:dist_fun}) and impurity velocity contributes significantly to the relative velocity in the deuterium-impurity collisions. Therefore additional corrections for the helium distribution function provide a larger effect than for the neon distribution function. 

It is interesting to mention that, for the deuterium, the main contribution into the shaping of the distribution function comes from the ion temperature gradient (figure \ref{fig:dist_fun}), which leads to the so-called "tail": overpopulated (with respect to the non-modified Maxwell function) high velocity phase space  and underpopulated low velocity phase space down the temperature gradient and opposite up the temperature gradient. In figure \ref{fig:dist_fun_outer} deuterium parallel flow velocity is positive (small), however the tail leads to a more effective momentum transfer to the impurity in the negative direction, due to a velocity dependence of the Coulomb collision cross-section. This phenomenon is the thermal force. It leads to the shift of the impurity distribution function. This shift leads to the friction force, which balances the thermal force. As a result, impurity parallel flow velocity is negative. This is the nature of the impurity thermal diffusion, such diffusion nature for electrons is described in \cite{braginskii1965transport}.

In figure \ref{fig:dist_fun_outer} one can see a negative part of the neon distribution. Such negative parts of distribution functions are present in this method or in \cite{braginskii1965transport}, however inside the area of theory applicability they are small and do not contribute to the result. In some places, where the ion temperature gradient is sufficiently strong (figure \ref{fig:dist_fun_outer}), the neon diffusive velocity $w_{\parallel Ne}$ is $\approx$60\% of the neon thermal velocity $\sqrt{T_i/m_{Ne}}$.  In such a case we are on the boundary of the applicability of linear Grad's-Zhdanov theory.
%or even cross it. 
Moreover, in the ITER simulations such distribution function approximation can be not sufficiently accurate, when it is applied for the ions heavier than neon, since their thermal velocity is even smaller. The applicability of this method for heavy impurities requires further investigations.

\section{Conclusions}
Improved analytical parallel kinetic coefficients are used for the helium and neon transport simulations for the ITER baseline scenario. New thermal and friction forces kinetic coefficients lead to weaker thermal diffusion, which drags impurity upstream. Significant impact on the helium transport is observed compared to the standard SOLPS-ITER approach. Relative helium concentration (separatrix-averaged) decreases by 30\%. For neon, changes are less pronounced. A simple leakage model is presented for illustration of the parallel impurity transport impact on the global impurity balance. Using linear Grad's-Zhdanov method ion distribution functions are analyzed for ITER Scrape-off layer predictive simulation for the first time. For helium this method is well applicable, whereas for neon ITER cases are on the boundary of applicability. 
%For the impurity heavier than neon the method could be out of area of applicability for ITER simulations.

\section*{Acknowledgments}

This work has been carried out within the framework of the EUROfusion Consortium and has received funding from the Euratom research and training programme 2014-2018 and 2019-2020 under grant agreement No 633053. This work was performed in part under the auspices of the ITER Scientist Fellow Network. The views and opinions expressed herein do not necessarily reflect those of the European Commission or of the ITER Organization.
\bibliographystyle{aip} 
\bibliography{refs}
\end{document}